\documentclass{article}

\usepackage{arxiv}

\usepackage[utf8]{inputenc} 
\usepackage[T1]{fontenc}    
\usepackage{hyperref}       
\usepackage{url}            
\usepackage{booktabs}       
\usepackage{amsfonts}       
\usepackage{amsmath}       
\usepackage{nicefrac}       
\usepackage{microtype}      
\usepackage{lipsum}
\usepackage{graphicx}
\graphicspath{ {./images/} }

\title{Deep Learning on Point Clouds for False Positive Reduction at Nodule Detection in Chest CT Scans.}

\author{
 Ivan Drokin \\
  Intellogic Limited Liability Company, \\
  territory of Skolkovo Innovation Center, \\
  121205, Moscow, Russia \\
  \texttt{ivan.drokin@botkin.ai} \\
   \And
 Elena Ericheva \\
  Intellogic Limited Liability Company, \\
  territory of Skolkovo Innovation Center, \\
  121205, Moscow, Russia \\
  \texttt{elena.ericheva@botkin.ai} \\
}

\begin{document}
\maketitle
\begin{abstract}
This paper focuses on a novel approach for false-positive reduction (FPR) of nodule candidates in Computer-aided detection (CADe) systems following the suspicious lesions detection stage. Contrary to typical decisions in medical image analysis, the proposed approach considers input data not as a 2D or 3D image, but rather as a point cloud, and uses deep learning models for point clouds. We discovered that point cloud models require less memory and are faster both in training and inference compared to traditional CNN 3D, they achieve better performance and do not impose restrictions on the size of the input image, i.e. no restrictions on the size of the nodule candidate. We propose an algorithm for transforming 3D CT scan data to point cloud. In some cases, the volume of the nodule candidate can be much smaller than the surrounding context, for example, in the case of subpleural localization of the nodule. Therefore, we developed an algorithm for sampling points from a point cloud constructed from a 3D image of the candidate region. The algorithm is able to guarantee the capture of both context and candidate information as part of the point cloud of the nodule candidate. We designed and set up an experiment in creating a dataset from an open LIDC-IDRI database for a feature of the FPR task, and  is herein described in detail. Data augmentation was applied both to avoid overfitting and as an upsampling method. Experiments were conducted with PointNet, PointNet++, and DGCNN. We show that the proposed approach outperforms baseline CNN 3D models and resulted in 85.98 FROC versus 77.26 FROC for baseline models. We compare our algorithm with published SOTA and demonstrate that even without significant modifications it works at the appropriate performance level on LUNA2016 and shows SOTA on LIDC-IDRI.
\end{abstract}


\section{Introduction}

Survival in lung cancer (over a period of 5 years) is approximately 18.1\%\footnote{2018 state of lung cancer report: https://www.naaccr.org/2018-state-lung-cancer-report/}. This rate is significantly lower than in other types of cancer owing to symptoms of this disease usually only appearing when the cancer is already at a late stage. Early-stage lung cancer (stage I) has a five-year survival rate of 60-75\%. A recent National Lung Screening Trial (NLST) study has shown that lung cancer mortality can be reduced by at least 20\%, using a high-risk annual screening program with low-dose computed tomography (CT) of the chest \cite{Introduction_2}. Computerized tools, especially image analysis and machine learning, are key factors for improving diagnostics, facilitating the identification of results that require treatment, and supporting the workflow of an expert \cite{Introduction_6}.

CAD systems for computed tomography (CT) typically involve two steps: suspicious lesions (pulmonary nodules candidates) detection and a false positive reduction stage. Although CAD systems have shown improvements in readability by radiologists \cite{Introduction_8,Introduction_9,Introduction_10}, a significant number of nodules remained undetected at a low rate of false-positive results, prohibiting the use of CAD in clinical practice. Classification tasks in the medical domain are often a normal vs pathology discrimination task. In this case, it is worth noting the normal class is extremely over-represented in a dataset. Furthermore, it has been shown \cite{Introduction_7} that nodules have a wide variation in shapes, sizes, and types (e.g., solid, sub-solid, calcined, pleural, etc.). Most normal training patterns are highly correlated due to the repetitive pattern of normal tissues in each image. 

Until recently, CAD systems were built using manually created functions and decision rules. With the advent of a new era of deep learning, the situation has changed. Now, to solve the detection task, CNN models are widely studied. Due to their specificity, CNNs can work efficiently with images and focus on candidate recognition \cite{Introduction_6}. However, due to architectural limitations, convolution neural networks can only function effectively with data of a strictly specified size. The selection of images of incorrect sizes can significantly reduce the effectiveness of the model for objects that are either too large or too small. In this work, we have proposed a novel approach for the false-positive reduction stage for CAD systems, based on point cloud networks. These networks require less memory for training and inference, show more stable results when working with multiple data sources and, most importantly, do not impose restrictions on the size of the input image - unlike CNN counterparts.

In some cases, the volume of the nodule candidate can be much smaller than the surrounding context, for example in the case of subpleural nodule localization. We have proposed an algorithm for transforming 3D CT scan data to point cloud.  

We have developed an algorithm for sampling points from a point cloud, constructed from a 3D image of the candidate region. The algorithm must be able to capture both context and candidate information as part of the point cloud of the nodule candidate.

The  goal  of  the  false-positive  reduction  task  is  to  recognize a true  pulmonary nodule from multiple plural candidates, which are received from the detection stage. Both point cloud sampling methods and FPR-model inputs strongly depend on segmentation masks suitable for nodule candidates. Performance evaluation of the FPR is both deeply connected with and dependent on the performance of a detector. In order to  ascertain the accuracy of the  FPR model performance (investigated separately from the detector) we designed and have described the full process of receiving a special artificial dataset for FPR training and evaluation.

This article consists of the following structure. In the \textit{\nameref{PrevWork}} section we have provided an overview of major works on lung cancer detection and lungs nodules evaluation.  \textit{\nameref{PCN}} provides an overview of major works and models on point clouds. Under \textit{\nameref{DnP}} we describe data processing for training both point-cloud-based and baseline models. In the section \textit{\nameref{PCsamplin}} we explain in detail a point cloud sampling policy and, in \textit{\nameref{AUG}}, an augmentation policy. All the experiments we conducted on LIDC-IDRI and LUNA2016 datasets are described under the  \textit{\nameref{EXP}} section with a comparison of current SOTA.

\section{Previous work}
\label{PrevWork}
Several articles \cite{Prev_1a,Prev_1b} deal with the main concepts of deep learning that are related to the analysis of medical images, and summarize more than 300 articles in this field. The authors consider the use of deep learning for image classification, object detection, segmentation, registration, and other tasks aimed at showing that deep learning methods have spread across the entire field of medical image analysis. The surveys identify problems in the successful application of deep learning for medical imaging tasks, and in addition illustrate specific solutions that solve or circumvent these problems. 

As previously explained, currently the conventional pipeline in the screening task for CAD consists of several stages - principally detection and cancer classification. A two-stage machine learning algorithm is a popular approach that can assess the risk of cancer associated with a CT scan \cite{Prev_12a,Prev_12b,Prev_14a,Prev_14b,Prev_13}. The first stage uses a nodule detector which identifies nodules contained in the scan. The second step is used to assess whether nodules are malignant or benign.

Methods to solve false positive reduction tasks separately from full CAD pipelines have been highly favored in recent times. A multicontext three‐dimensional residual convolutional neural network (3D Res‐CNN) was proposed in \cite{Prev_15a} to reduce false-positive nodules. 3D Res-CNN uses a shortcut connection to realize the residual structure. It uses two network scales to adapt the variation of pulmonary nodule size, rather than an unreferenced function which carries out identity mapping. 
The proposed deep 3D residual CNN in \cite{Prev_15b} is much more complex than the traditional 3D CNNs used in medical image processing. A spatial pooling and cropping (SPC) layer to extract multi-level contextual information of CT data was designed and added to this 3D Res-CNN. An online hard sample selection strategy is applied in the training process to allow the network to better adapt hard samples (e.g. nodules with irregular shapes). 
The experiments on the LUNA16 dataset confirm that their method is robust and that the proposed spatial pooling and cropping layer helps increase prediction accuracy. The method presented in \cite{Prev_15c} is based on structural relationship analysis between nodule candidates and vessels, and the modified surface normal overlap descriptor. This approach is based on an assumption that a large number of false nodules attached to vessels can be removed by the relationship analysis between nodule candidates and their attached tissues. At the same time, low-contrast nonsolid nodules are separated from the candidates with the modified surface normal overlap descriptor. 
The algorithm proposed in \cite{Prev_15d} segments lungs and nodules through a combination of 2D and 3D region growing, thresholding and morphological operations. To reduce the number of false positives, a rule-based classifier is used to eliminate obvious non-nodules, followed by a multi-view Convolutional Neural Network. This solution achieves a good tradeoff between efficiency and effectivity, and saves significant computation time. The proposed multi-view 2D network can detect nodules that are isolated, linked to a vessel or attached to the lung wall. 
The CNN from \cite{Prev_15e} is fed with nodule candidates obtained by combining three candidate detectors specifically designed for solid, subsolid, and large nodules. For each candidate, a set of 2D patches from differently oriented planes is extracted. The proposed architecture comprises multiple streams of 2D CNN, from which the outputs are combined using a dedicated fusion method to acquire the final classification. 
An evaluation was performed on independent datasets from the LUNA16\footnote{LUNA16 challenge homepage: https://luna16.grand-challenge.org/} and ANODE09\footnote{ANODE09 challenge homepage: https://anode09.grand-challenge.org/} challenges and DLCST\cite{Prev_16}.

Recently, several successful works using point clouds in medical image analysis have appeared. In \cite{Experiments_5}, a segmentation of teeth was presented. In \cite{Experiments_5b}, segmentation refinement with false positive reduction by point clouds was proposed. In \cite{Experiments_5a} the authors use point clouds for vertebra shape analysis.

\section{Point Clouds}
\label{PCN}
A point cloud is represented as a set of 3D points $\{P_i | i = 1, ..., n\}$, where each point $P_i$ is a vector of its $(x, y, z)$ coordinate in addition to extra feature channels. In semantic segmentation, the input is a single object for part region segmentation, or a sub-volume from a 3D scene for object region segmentation. The input is a subset of points from an Euclidean space and satisfies several properties. Unlike pixel arrays, it is unordered in images or voxel arrays in volumetric grids. The points are from a space with a distance metric. This means that points are not isolated, and neighboring points form a meaningful subset. As a geometric object, the learned representation of the point set should be invariant to certain transformations. A neural network, named PointNet, was shown in \cite{Experiments_2}. It directly works with point clouds, and fully respects the permutation invariance of points in the input. Therefore, the model is able to capture local structures from nearby points, and the combinatorial interactions among local structures. The network provides a unified architecture for applications, ranging from object classification and part segmentation to scene semantic parsing. Though simple, PointNet is highly efficient and effective. Empirically, it shows strong performance on par with or even better than state of the art. 

In \cite{Experiments_3}, the authors expand further on the idea with added hierarchical structures. These allow PointNet to capture local structures induced by the metric space points live in, as well as increasing its ability to recognize fine-grained patterns and generalizability to complex scenes. In that study a discrete metric space $X = (P_i, d)$ is considered as input. The distance metric $d$ defines local neighborhoods that may exhibit different properties. For example, the density and other attributes of points $ \{P_i | i = 1, ..., n\}$ may not be uniform across different locations. Point sets are usually sampled with varying densities, which results in greatly decreased performance for networks trained on uniform densities. While PointNet uses a single max pooling operation to aggregate the whole point set, PointNet++ builds a hierarchical grouping of points and progressively abstracts larger and larger local regions along the hierarchy. By exploiting metric space distances, proposed hierarchical PointNet++ is able to learn local features with increasing contextual scales by applying PointNet recursively on a nested partitioning of the input point set. Experiments show that PointNet++ attentively combines features from multiple scales and is able to learn deep point-set features efficiently and robustly.

Finally, in \cite{Experiments_4}, the authors propose a module dubbed EdgeConv suitable for CNN-based high-level tasks on point clouds, including classification and segmentation. EdgeConv is differentiable and can be plugged into existing architectures. Instead of working on individual points like PointNet, however, EdgeConv exploits local geometric structures by constructing a local neighborhood graph and applying convolution-like operations on the edges, connecting neighboring pairs of points (as in graph neural networks). Proximity in feature space differs from proximity in the input, leading to nonlocal diffusion of information throughout the point cloud. Unlike graph CNNs, an EdgeConv graph is not fixed but instead is dynamically updated after each layer of the network. The set of k-nearest neighbors of a point changes from layer to layer of the network and is computed from the sequence of embeddings. In multi-layer systems with an affinity in feature space, EdgeConv captures semantic characteristics over potentially long distances in the original embedding. Extensive evaluation reveals that EdgeConv captures and exploits fine-grained geometric properties of point clouds, compared to existing modules which operate largely in extrinsic space or treat each point independently. 

\section{Data and Processing}
\label{DnP}

\begin{figure}[htbp]
\centering
  \includegraphics[width=1.0\linewidth]{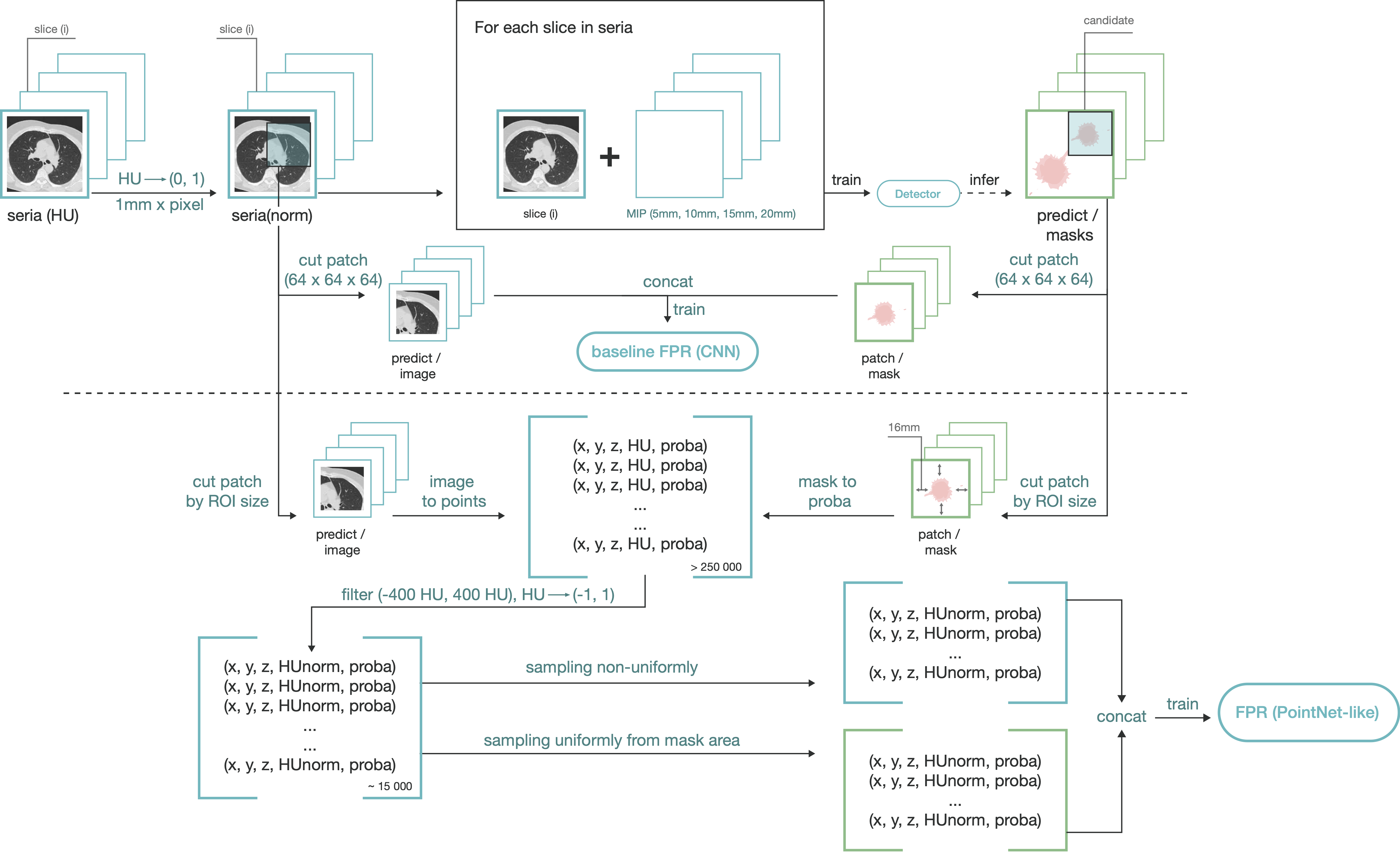}
  \caption{Full pipeline of data processing}
  \label{fig:figDataPipeline}
\end{figure}

It is difficult to directly and objectively compare different CAD systems. In \cite{Data_4}, an evaluation system was proposed for the automatic detection of nodules on CT images. A large dataset, containing 888 CT scans with annotations from the open LIDC-IDRI database\footnote{LIDC-IDRI Dataset: https://wiki.cancerimagingarchive.net/display/Public/LIDC-IDRI}, is available from the NCI Cancer Imaging Archive\footnote{TCIA Collections: https://www.cancerimagingarchive.net}. A detailed description of the data set is given in \cite{Data_7}. This includes a description of the range of possible patients and disease manifestations, as well as the data labels reception process, and an analysis of their ability to support the performance of this method. 
Any nodule noted by at least one radiologist was considered as positive ground truth. This increased the number of positive ROI in our dataset. In general this led to a decrease in recall at low levels of false positives. Additionally, we took into account all nodules $\leq 3$ as a false positive. Our reasoning for choosing such a markup is the goal of constructing a pipeline for detecting nodes with maximum recall, including controversial cases. In our opinion, such borderline situations should be detected by an algorithm, and a doctor should have to make a final decision regarding this region. It is in controversial and non-obvious cases that the new CAD systems should be of help to the doctor. Since different series in the dataset can have different slice thickness parameters, and thus different distances between slices, we resampled series space to 1 mm per voxel side. The full data preprocessing pipeline is shown in Figure \ref{fig:figDataPipeline}.

We described dataset collection for the false positive reduction task as a subsection \textit{\nameref{FPRDataset}}. We have also explained preprocessing algorithms for each model in relevant subsections: for detector in the \textit{\nameref{DDP}} section, for baseline FPR in the \textit{\nameref{BFPR}} section, for point cloud-based in the \textit{\nameref{preproc}} section.

\subsection{Collecting a dataset for FPR task}
\label{FPRDataset}
\begin{figure}[htbp]
\centering
  \includegraphics[width=0.8\linewidth]{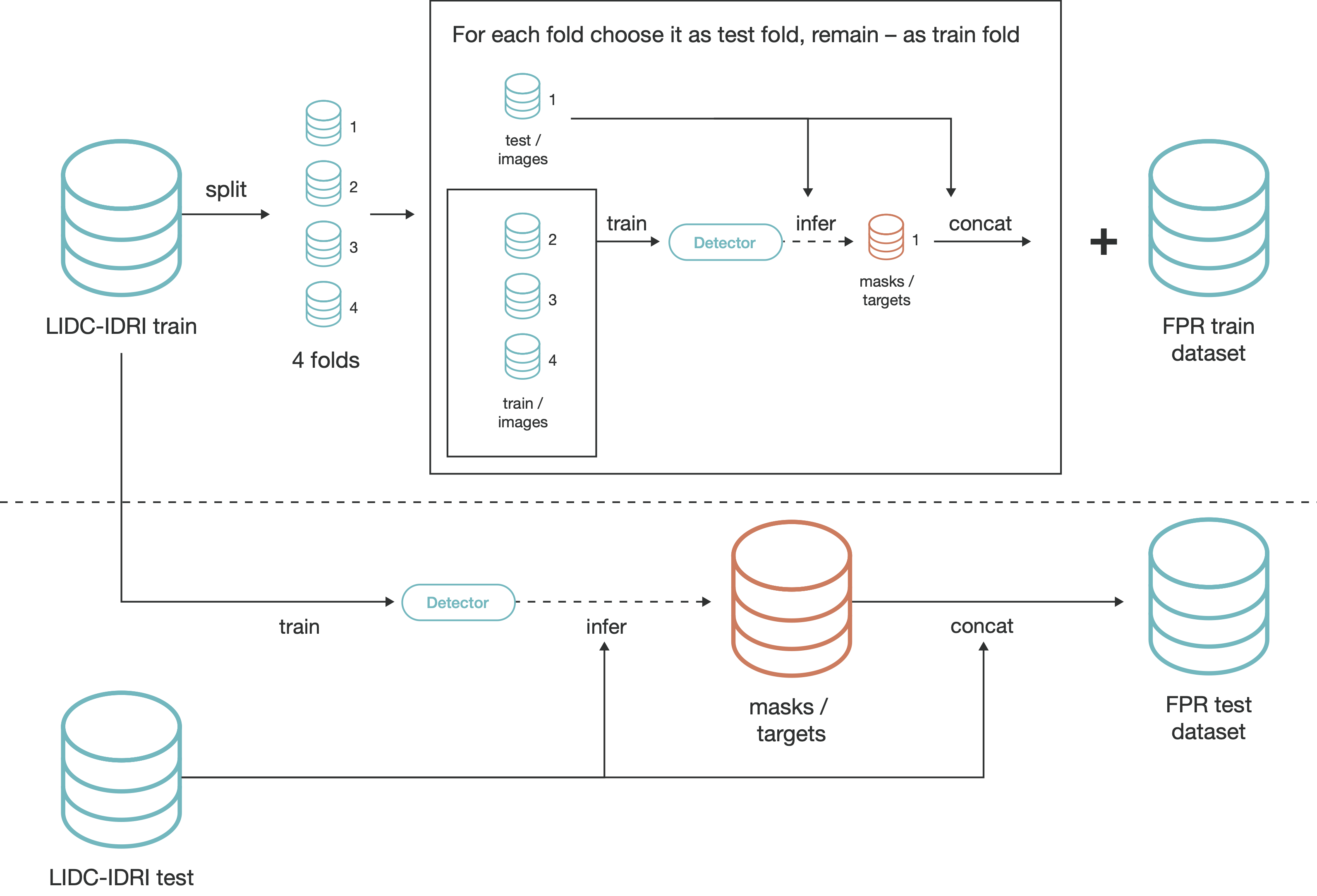}
  \caption{Full scheme of the artificial dataset for FPR task}
  \label{fig:FPRDataset}
\end{figure}

Since both our point cloud sampling method and FPR-model input strongly depend on segmentation masks suitable for nodule candidates, we have used the LIDC-IDRI dataset as a source for setting up training and testing data for the FPR task. A simple scheme is presented in Figure \ref{fig:FPRDataset}.
We have used DenseNet121-TIRAMISU \cite{TrainingData_1} as a detector for nodule candidates, as well as for train and test datasets for FPR task collection. Firstly, we split LIDC-IDRI datasets into train and test datasets with a 75\%-25\% ratio. For the creation of the FPR-train dataset we applied 4-fold cross-validation. We split the LIDC-IDRI train dataset into 4 subsets, trained a detector on 3 of these subsets, then inferred the detector on the remaining subset. The data received from the inference phase, including predicted segmentation masks and probabilities, is then appended to the training dataset for the FPR task. We repeated this detector train-infer loop 4 times, consistently considering each of the 4 folds from LIDC-IDRI as a subset to infer. To create the FPR-test dataset we trained the detector on the full LIDC-IDRI train dataset, then inferred the detector on the LIDC-IDRI test dataset. The data received from the inference is then used as a test dataset for the FPR task.

\subsection{Data Preprocessing for detector}
\label{DDP}
The images are presented on the Hounsfield scale\footnote{https://en.wikipedia.org/wiki/Hounsfield\_scale}. Voxel intensities are limited to an interval from -1000 to 400 HU and normalized in the range from 0 to 1 (see Equation \ref{eq:HUnorm}). 
    \begin{equation}
    x_{norm} = \frac{x_{HU} + 1000}{1400}\label{eq:HUnorm}
    \end{equation}
We used as a detector input a full 2D slice concatenated with 4 Maximum Intensity Projection (MIP) images \cite{Data_12} per sample. MIP images are used by radiologists alongside a complete CT exam in order to improve the detection of pulmonary nodules, especially small nodules. MIP images consist of the superposition of maximum grey values at each coordinate from a stack of consecutive slices. Such a combined image shows morphological structures of isolated nodules and continuous vessels. Experimental results in \cite{Data_12} showed that utilizing MIP images can increase sensitivity and lower the number of false positives, which demonstrates the effectiveness and significance of the proposed MIP-based CNNs framework for automatic pulmonary nodule detection in CT scans. We took MIP images of different slab thicknesses of 5mm, 10mm, 15mm and 20mm and 1mm axial section slices as input.

\subsection{Data Preprocessing for baseline CNN-like FPR model}
\label{BFPR}
For each predicted region of interest (ROI) from the detector network, we cut a patch sized $64\times64\times64$, the center of which is equal to the center of the segmented nodule area provided by the detector. The patch, normalized with Equation \ref{eq:HUnorm}, is considered as an input to the network.

\subsection{Data Preprocessing for PointNet-like FPR models}
\label{preproc}
For each predicted region of interest (ROI) from the detector network, we form a bounding box with 16 mm padding. Bounding boxes here can be any size, matching the binarized segmenting mask obtained from a detector. Thus we can provide context for the candidate and the surrounding neighborhood, with different lung tissues like vessels, pleura, bronchi, etc.
To reduce the number of points and exclude the most non-informative points from consideration we selected points only with a HU value from -400 to 400 from this bounding box. This allows us to catch the vast majority of all nodule types and, at the same time, reduces the number of points from 250,000 to 15,000 on average. 
The input point consists of several features: coordinates of a point, its HU density value, and the probability that it will be predicted by the detector. As a normalization step, we centered coordinates over the center of the segmented nodule area provided by the detector, and in addition translated HU values to a (-1,1) interval. The full pipeline scheme of data processing is presented in Figure \ref{fig:figDataPipeline}.

\section{Point Cloud Sampling}
\label{PCsamplin}
For the basis of our proposed point cloud sampling method, we used the method described in \cite{Experiments_5}. 
A naive resampling approach can invoke the loss of important information and is highly application-dependent. Since 2016, several studies have investigated point cloud analysis \cite{Experiments_2,Experiments_3} but still applied a uniform resampling for fixing the number of points. Such an approach does not preserve the finer details of data, which is important for the segmentation tasks. Furthermore, it also ignores the existing strong dependency between the label of each point and its location in the point cloud. In \cite{Experiments_5}, the authors propose a unique non-uniform resampling mechanism that facilitates the network on a fixed-size resampled point cloud, which contains different levels of the spatial resolution, involving both local, fine details and the global shape structure. It is based on the Monte Carlo sampling technique and a Radial Basis Function ($RBF$, Equation \ref{eq:RBF})
    \begin{equation}\label{eq:RBF}
    RBF(x_i, x_{center}) = 
    exp(- \frac{||x_i - x_{center}||^2}{2 \sigma^2})
    \end{equation}
Let $X \in R^{N*3}$ be a full point cloud with attributes obtained as described in Subsection \ref{preproc}. $x_{center} \in X$ - is a randomly chosen point. Geometrical similarity (spatial distance) to the point $x_{center}$ can then be measured with a weighted distance metric. By resampling, we aim to choose a subset out of $X$ with $M$ points ($M < N$) that has a dense sampling around the $x_{center}$ and a sparse sampling for farther locations. By randomly drawing (with a replacement) a point $x_i$ from the set $X$, we accept to insert such a point into the target subset only if $RBF > \tau$. The variable $\tau$ is a random number from a uniform distribution within the unity interval according to the Monte Carlo technique. $\sigma$ is a parameter that controls the bandwidth (compactness) of the kernel and depends on candidate radius $r$.

We introduced two major differences from this method. First, we change the $\sigma$ parameter that controls the bandwidth (compactness) of the kernel and depends on the candidate radius $r$. We choose $\sigma = r/2$. Second, we introduced additional sampling from the nodule area. It should be pointed out that there is a problem of sampling small subpleural candidates with a diameter $<5$ mm and an overwhelming surrounding context. The probability of the candidate’s point inclusion to the sampled point cloud decreases with the reduction of the candidate’s size. To deal with this, we sample additionally and uniformly only from the area corresponding to the segmenting mask provided by the detector. Thus we can guarantee the appearance of candidate points in the sampled point cloud regardless of the candidate's location or size. 

The basic intuition that underlies the proposed sampling method is as follows: the farther the point is from the center of the candidate, the less likely it will be selected in the sampling process. To control this, we introduce sigma as half the candidate radius. Obviously, the farther the point is from the candidate, the less likely it is to be classified as a nodule / non-nodule. For example, even if the detector incorrectly selected a part of the vessel as a nodule, the points of interest that are closest to the candidate's points are of greatest interest. This will allow the algorithm to decide that it is a vessel (or bronchus), and not a nodule. Whether or not a point belongs to the candidate is determined by the order of the detector in binarizing its mask. Not only does all of this guarantee a context capture regardless of the size of the original candidate, but uniform sampling also provides us with all the necessary information about the region of interest.

\section{Augmentation techniques}
\label{AUG}
We are dealing with an imbalance between false positives and nodules. In order to avoid overfitting, augmentations are performed at an image level as well as at a point cloud level. At an image level, we apply Gaussian noise blur and Hounsfield units shift.
    
    \begin{equation}
    \begin{split}
    x_{HU}' = x_{HU} + m_{ijk}*n_{ijk}\label{eq:maskednoise}\\
    m_{ijk} \sim Poisson(\lambda), n_{ijk} \sim N(0, \sigma) \nonumber
    \end{split}
    \end{equation}
    
Gaussian noise is added not to a whole slice but rather in accordance with the random generated mask. In Equation \ref{eq:maskednoise}, $x_{HU}$ is an input image in Hounsfield units, and an addendum is a matrix with its shape equal to the image shape. Binary mask $m_{ijk}$ is multiplied with values $n_{ijk}$ sampled from Gauss distribution $N(0, \sigma)$). We have used 0.2 for blur appearance probability and the interval [0.2, 0.8] for the Gaussian filter coefficient. The Gaussian filter coefficient alpha is distributed uniformly. 
    \begin{equation}
    \begin{split}
    \label{eq:coordinationshift}
    {x_{inp}}' = x_{inp}*(I + 
    \begin{pmatrix}
    N(0,\sigma) & 0 & \cdots & 0 \\
    0 & N(0,\sigma) & \cdots & 0 \\
    \vdots  & \vdots  & \ddots & \vdots  \\
    0 & 0 & \cdots & N(0,\sigma) 
    \end{pmatrix})
    \end{split}
    \end{equation}
At the point cloud level we added a rotation (at a randomly chosen degree in the transverse planes) and a random constant coordination shift. This shift presents compression and extension (Equation \ref{eq:coordinationshift}, where $x_{inp}$ - model's input point cloud,  $I$ - is an identity-matrix).  Balanced random selection from the data set was applied.  The augmentation technique is applied only to the training dataset; the test data is utilised as is. The augmentation technique is applied both for CNN-like baselines as well as for PointNet-like FPR models.

\section{Experiments}
\label{EXP}
\subsection{Experiments on LIDC-IDRI}
\begin{figure}[htbp]
\centering
\includegraphics[width=.4\linewidth]{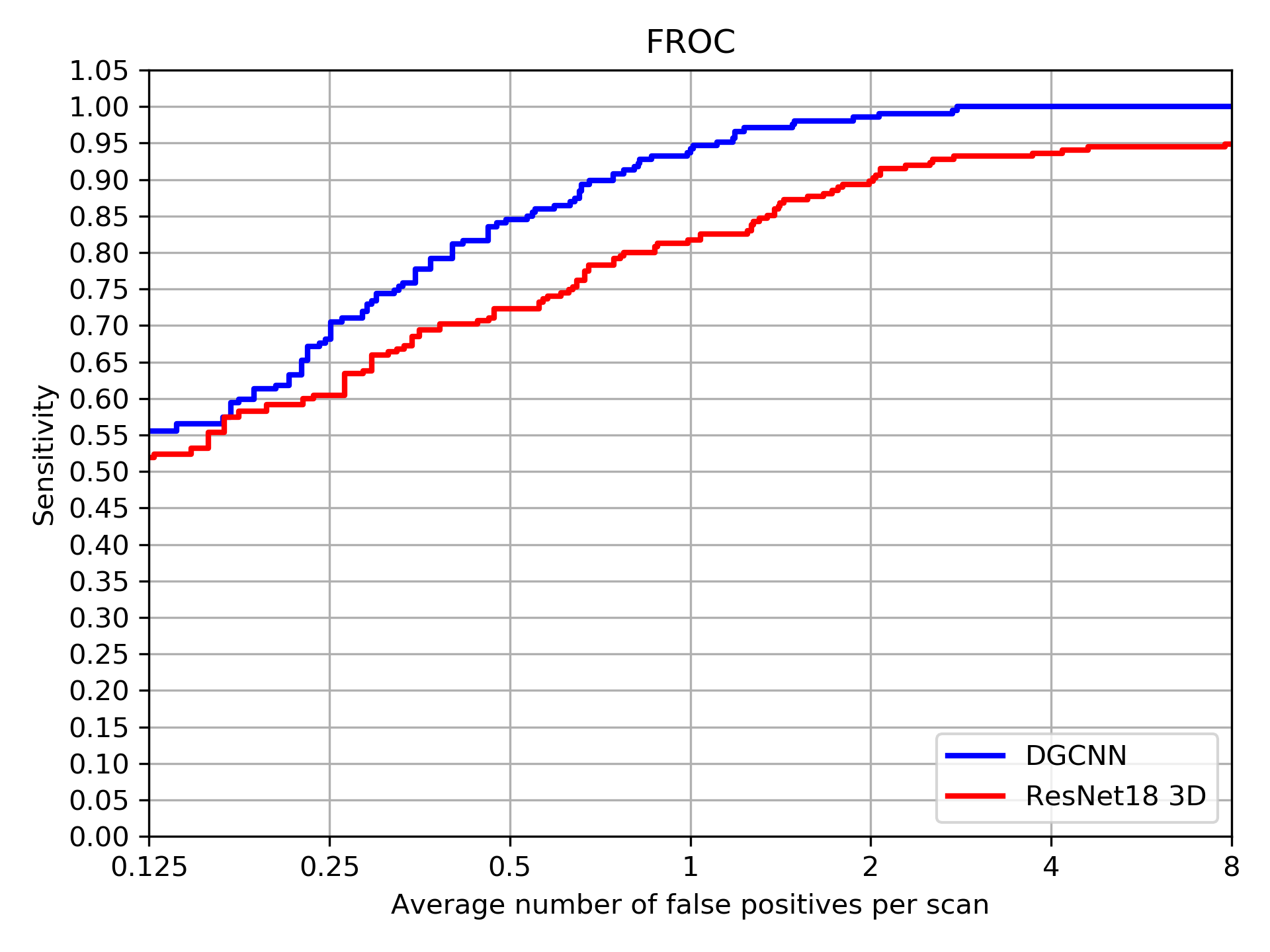}
\caption{FROC. red: for baseline CNN 3D model and blue: for DGCNN model}
\label{fig:test}
\end{figure}

The goal of the false-positive reduction task is to recognize true pulmonary nodules from a variety of candidates, which are identified at the first step of pulmonary nodule candidate detection. Here we investigated FPR performance separately from the detector results. We considered the simplest and most common architecture of a detector model and used it only as a sampler for creating the FPR train dataset. Performance evaluation of the FPR is deeply connected with and dependent on the performance of a detector. That is why we examined FPR performance scoring only on an artificially received dataset, as described in Section \ref{FPRDataset} and in Figure \ref{fig:FPRDataset} 

\begin{table}[htbp]
\centering
\addtolength{\tabcolsep}{-0.5pt}
  \caption{Results of the experiments on LIDC-IDRI. Sensitivity per FP level per exam.}
  \setlength{\tabcolsep}{1pt}
  \begin{tabular}{|c|c|c|c|c|c|c|c|c|}
  \hline
  \bfseries Experiment & \bfseries 0.125 FP & \bfseries 0.25 FP & \bfseries 0.5 FP & \bfseries 1 FP & \bfseries 2 FP & \bfseries 4 FP & \bfseries 8 FP & \bfseries Mean Sens\\\hline
  Li \cite{Experiments_7} & 0.600 & 0.674 & 0.751 & 0.824 & 0.850 & 0.853 & 0.859 & 0.773\\
  Liao \cite{Experiments_6} & 0.662 & 0.746 & 0.815 & 0.864 & 0.902 & 0.918 & 0.932 & 0.834\\\hline
  Baseline & 0.514 & 0.600 & 0.710 & 0.812 & 0.893 & 0.931 & 0.944 & 0.772\\
  Pointnet & 0.433 & 0.560 & 0.693 & 0.835 & 0.946 & 0.977 & 0.982 & 0.775\\
  Pointnet w/aug & 0.438 & 0.607 & 0.698 & 0.844 & 0.949 & 0.990 & 0.990 & 0.788\\
  Pointnet++ & 0.360 & 0.502 & 0.640 & 0.776 & 0.922 & 0.972 & 0.995 & 0.738\\
  Pointnet++ w/aug & 0.356 & 0.502 & 0.657 & 0.789 & 0.922 & 0.990 & 0.995 & 0.744\\
  DGCNN & 0.497 & 0.628 & 0.758 & 0.904 & 0.969 & 0.994 & 0.994 & 0.821\\
  DGCNN w/aug & \textbf{0.545} & \textbf{0.679} & \textbf{0.842} & \textbf{0.971} & \textbf{0.990} & \textbf{0.995} & \textbf{0.995} & \textbf{0.859}\\\hline
  \end{tabular}
\label{tab:test}
\end{table}

We compared several architectures that work with point cloud data and baseline CNN. As a baseline we considered a ResNet3D model \cite{Experiments_1}, trained on the same dataset (See Appendix \ref{app:app1}). We examined 3 PointNet based models: PointNet \cite{Experiments_2}, PointNet++ \cite{Experiments_3} and DGCNN \cite{Experiments_4}. We used FROC \cite{froc_art} as the model's quality criterion, as a natural metric for nodule detection systems. During experiments, we performed several runs for selected architectures and evaluated model performance with and without augmentations during the training procedure. We used an ADAM \cite{adam} optimizer with a starting learning rate of 0.001, and trained models during 70 epochs, decreasing the learning rate twice every 10 epochs. We achieved the best FROC of 85.98 using the DGCNN model with augmentation. This result outperforms the 77.26 FROC of the baseline model. Results for all the experiments are shown in Table \ref{tab:test}. In this table, Sens./0.125 FP means that models show sensitivity equal to a number from the corresponding table cell at a mean 0.125 False Positives per scan. In the table we have also provided recent published results(\cite{Experiments_6}, \cite{Experiments_7}). Our CNN baseline and our PointNet-like approaches demonstrated a performance matching the level of \cite{Experiments_7}. At the same time, the best DGCNN with augmentation outperformed \cite{Experiments_6}. Examples of point cloud samples are provided in Appendix \ref{app:app2}.

\begin{figure}[htbp]
\centering
\includegraphics[width=.4\linewidth]{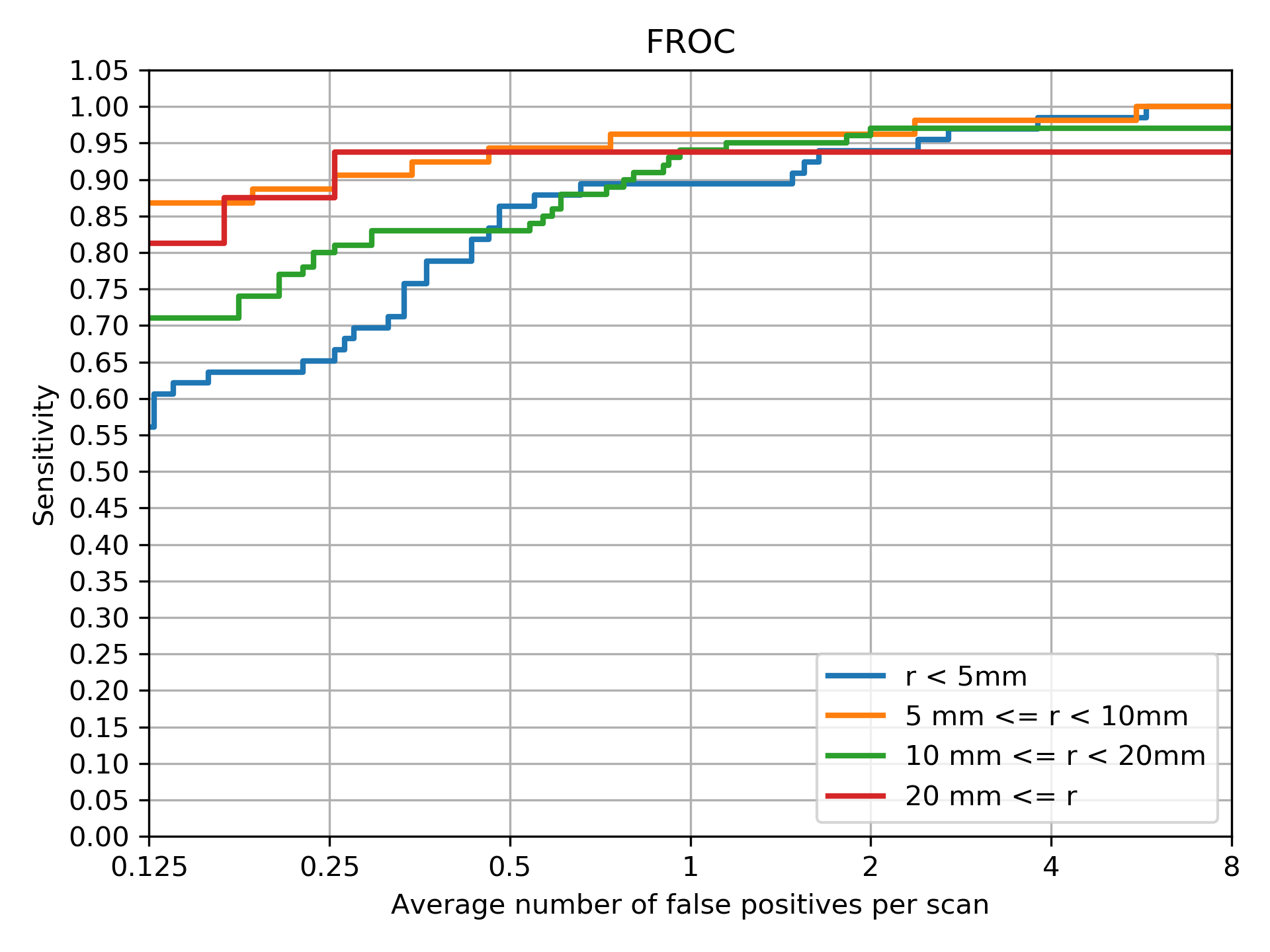}
\centering
\includegraphics[width=.4\linewidth]{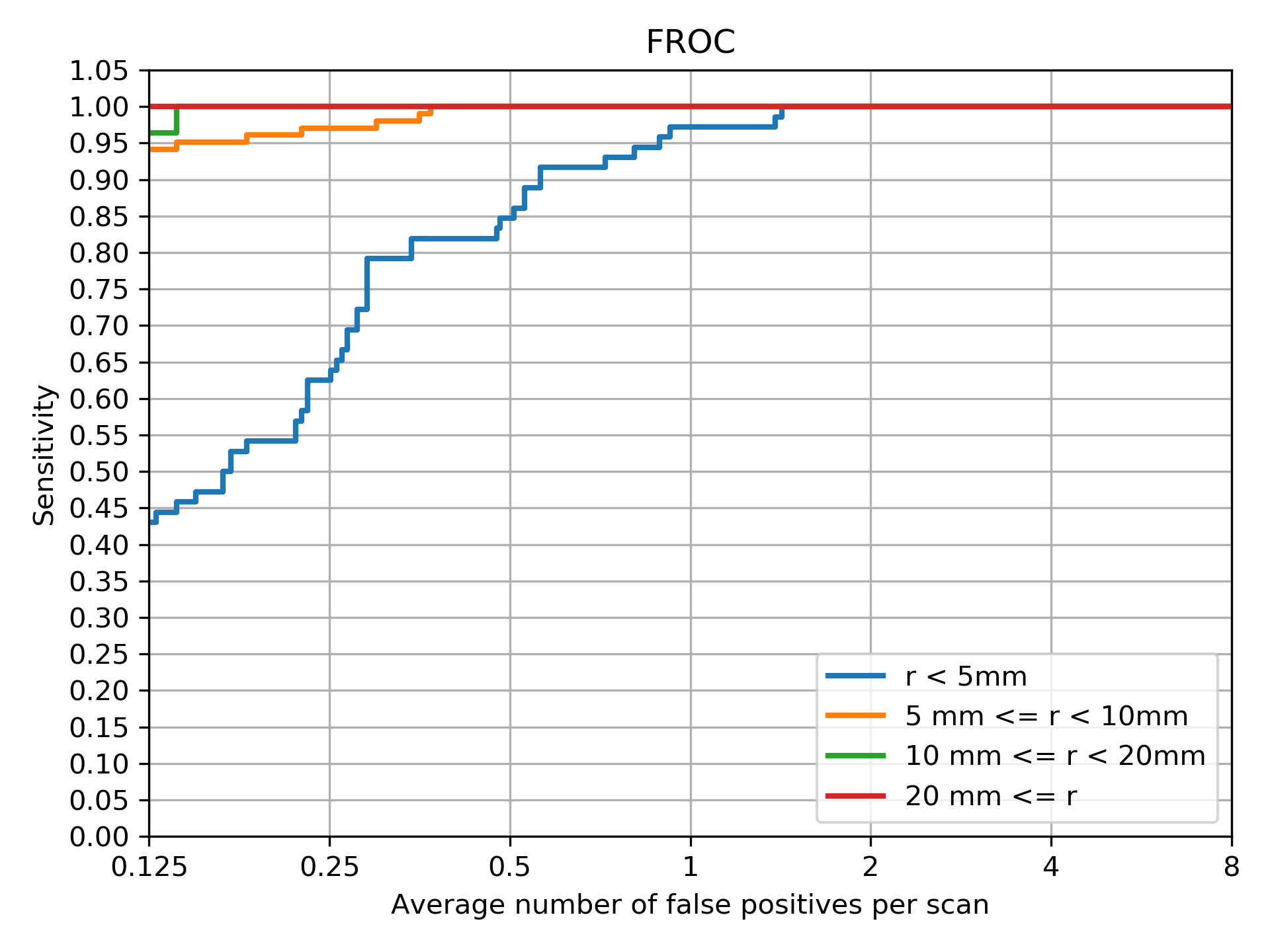}
\caption{FROC. left: for baseline CNN 3D model and right: for DGCNN model}
\label{fig:test1}
\end{figure}

\begin{table}[htbp]
\centering
\addtolength{\tabcolsep}{-0.5pt}
  {\caption{Results of the experiments on LIDC-IDRI. Sensitivity per FP level per exam. Comparison of the performance of FPR models with various point sampling methods.}}
  {\setlength{\tabcolsep}{1pt}
  \begin{tabular}{|c|c|c|c|c|c|c|c|c|}
  \hline
  \bfseries{\begin{tabular}{@{}c@{}} Experiments \\ DGCNN w/aug\end{tabular}} & \bfseries 0.125 FP & \bfseries 0.25 FP & \bfseries 0.5 FP & \bfseries 1 FP & \bfseries 2 FP & \bfseries 4 FP & \bfseries 8 FP & \bfseries Mean Sens\\\hline
  \begin{tabular}{@{}c@{}}w/uniform \\ sampling\end{tabular} & 0.36 & 0.52 & 0.655 & 0.78 & 0.902 & 0.969 & 0.99 & 0.7394\\
  \begin{tabular}{@{}c@{}}w/(Section \ref{PCsamplin}) \\ sampling \end{tabular} & \textbf{0.545} & \textbf{0.679} & \textbf{0.842} & \textbf{0.971} & \textbf{0.990} & \textbf{0.995} & \textbf{0.995} & \textbf{0.859} \\\hline
  \end{tabular}}
\label{tab:test3}
\end{table}

We added the FROC pictures depending on the nodule size. The results  show that in a case with large nodules there is no performance degradation. Furthermore, PointNet-like models performed better in such cases compared to a CNN approach (Figure \ref{fig:test1}).

To show the effect of the selected sampling method on the performance of the model, we conducted additional experiments with a uniform method of sampling. Table \ref{tab:test3} shows a significant increase in the FROC values for each FP level in cases with proposed sampling, compared to uniform sampling.

\subsection{Experiments on LUNA2016 nodule detector track. Benchmark}
\begin{table}[htbp]
\centering
\addtolength{\tabcolsep}{-0.5pt}
  {\caption{Results of the experiments on LUNA nodule detection track. Sensitivity per FP level per exam}}
  {
  \setlength{\tabcolsep}{2pt}
  \begin{tabular}{|c|c|c|c|c|c|c|c|c|}
  \hline
  \bfseries Experiment & \bfseries 0.125 FP & \bfseries 0.25 FP & \bfseries 0.5 FP & \bfseries 1 FP & \bfseries 2 FP & \bfseries 4 FP & \bfseries 8 FP & \bfseries Mean Sens\\\hline
  Gupta \cite{Experiments_8} & 0.531 & 0.629 & 0.790 & 0.835 & 0.843 & 0.848 & 0.856 & 0.763\\
  Hamidian \cite{Experiments_9} & 0.583 & 0.687 & 0.742 & 0.828 & 0.886 & 0.918 & 0.933 & 0.797\\
  Xie \cite{Experiments_10} & 0.734 & 0.744 & 0.763 & 0.796 & 0.824 & 0.832 & 0.834 & 0.790\\
  Liao \cite{Experiments_6} & 0.662 & 0.746 & 0.815 & 0.864 & 0.902 & 0.918 & 0.932 & 0.834\\
  Dou \cite{Experiments_11} & 0.659 & 0.745 & 0.819 & 0.865 & 0.906 & 0.933 & 0.946 & 0.839\\
  Zhu \cite{Experiments_12} & 0.692 & 0.769 & 0.824 & 0.865 & 0.893 & 0.917 & 0.933 & 0.842\\
  Li \cite{Experiments_7} & 0.739 & 0.803 & 0.858 & 0.888 & 0.907 & 0.916 & 0.920 & 0.862\\
  Wang \cite{Experiments_13} & 0.676 & 0.776 & 0.879 & 0.949 & 0.958 & 0.958 & 0.958 & 0.878\\
  Ding \cite{Experiments_14} & 0.748 & 0.853 & 0.887 & 0.922 & 0.938 & 0.944 & 0.946 & 0.891\\
  Khosravan \cite{Experiments_15} & 0.709 & 0.836 & 0.921 & 0.953 & 0.953 & 0.953 & 0.953 & 0.897\\
  Ozdemir \cite{Experiments_16} & 0.832 & 0.879 & 0.920 & 0.942 & 0.951 & 0.959 & 0.964 & 0.921\\
  Cao \cite{Experiments_17} & 0.848 & 0.899 & 0.925 & 0.936 & 0.949 & 0.957 & 0.960 & 0.925 \\\hline

  Baseline & 0.686 & 0.811 & 0.840 & 0.929 & 0.972 & 0.972 & 0.972 & 0.883\\
  DGCNN w/aug & \textbf{0.725} & \textbf{0.832} & \textbf{0.901} & \textbf{0.933} & \textbf{0.945} & \textbf{0.945} & \textbf{0.945} & \textbf{0.8894} \\\hline
  \end{tabular}}
\label{tab:test2}
\end{table}

The generally accepted benchmark for the FPR task in nodule detection is the LUNA2016 FPR track competition. We believe that a comparison of our results with the existing benchmark is important for a more transparent reflection on the results obtained and the performance of the model. However, it is also necessary to note that we are unable to use the LUNA2016 FPR track data as a benchmark for the point-cloud-based model. This is because the data on this track does not have the necessary markup. Our FPR model inputs strongly depend on segmentation masks suitable for nodule candidates(see Section \ref{FPRDataset}).

Nevertheless, we are able to evaluate the entire pipeline on the LUNA2016 nodule detection track. We are aware of the heavy dependence of the whole pipeline evaluation on the performance of the detector. FNs cannot be recovered by the FPR model. A weak detector can create many FNs. Furthermore we want to note that the study of the performance of the detector is beyond the scope of this work. Nevertheless, for a more transparent estimation of the FPR models, we briefly present the results of the detector performance of the LUNA2016 nodule detection track: mean sensitivity 0.4132, max recall 0.9776 achieved on 43 FP per case.

Table \ref{tab:test2} shows that both the CNN baseline and the PointNet approaches in conjunction with a weak detector have a performance comparable to the level of recent published results.

\section{Ablation study}
As part of the formulation of the general objective of this work — testing the hypothesis that lung nodules can be effectively represented as a point cloud  — we decided that it would be interesting to include the results of an ablation study. First, to clarify the problem: patch classification refers to the classification of objects presented on these patches - the nodule is either represented there or not. Common false-positive detections include pieces of bronchi, blood vessels, fibrosis, etc. In the task of separating these objects from the nodules, information about their shape is critical. As a result, coordinates are needed as data for representation. The coordinates contain information about the shape of the object represented on the patch. Coordinates "structure" points in a point cloud. Speaking more formally, we propose to present objects not as a 3D image, but as a list of points with their own characteristics. The most obvious characteristic of a point is their coordinates $(x, y, z)$. We can expand this list by adding information about the radiological density ($HU$) and probability of the detector ($p$). Thus, each example is represented as a list ${(x_i, y_i, z_i, hu_i, p_i)} i = 1 ... n$, where $n$ is the number of points in the example. This data is already used by the model as an input. The results presented in Table \ref{tab:test4} confirm our assumptions on the importance of $HU$ and $p$ in the context of solving the FPR problem based on PointNet-like models, and presenting samples in the point cloud. Results of such an experiment show that coordinate information for each point is essential for successful nodule/non-nodule classification. This result is consistent with our hypotheses and assumptions.

\begin{table}[htbp]
\centering
\addtolength{\tabcolsep}{-0.5pt}
  {\caption{Results of the experiments on LIDC-IDRI. Sensitivity per FP level per exam. Comparison of the performance of FPR models with various point point representation in the Point Cloud.}}
  {\setlength{\tabcolsep}{1pt}
  \begin{tabular}{|c|c|c|c|c|c|c|c|c|}
  \hline
  \bfseries Experiment & \bfseries 0.125 FP & \bfseries 0.25 FP & \bfseries 0.5 FP & \bfseries 1 FP & \bfseries 2 FP & \bfseries 4 FP & \bfseries 8 FP & \bfseries Mean Sens\\\hline
  Li \cite{Experiments_7} & 0.600 & 0.674 & 0.751 & 0.824 & 0.850 & 0.853 & 0.859 & 0.773\\
  Liao \cite{Experiments_6} & 0.662 & 0.746 & 0.815 & 0.864 & 0.902 & 0.918 & 0.932 & 0.834\\\hline
  Baseline & 0.514 & 0.600 & 0.710 & 0.812 & 0.893 & 0.931 & 0.944 & 0.772\\
  DGCNN (xyz + HU + p) & \textbf{0.545} & \textbf{0.679} & \textbf{0.842} & \textbf{0.971} & \textbf{0.990} & \textbf{0.995} & \textbf{0.995} & \textbf{0.859}\\\hline
  DGCNN (xyz + p) & 0.382 & 0.577 & 0.756 & 0.902 & 0.959 & 0.979 & 0.992 & 0.792\\
  DGCNN (xyz + HU) & 0.362 & 0.467 & 0.630 & 0.764 & 0.862 & 0.979 & 0.995 & 0.723\\
  DGCNN (xyz) & 0.362 & 0.451 & 0.528 & 0.646 & 0.829 & 0.947 & 0.995 & 0.680\\
  DGCNN (HU + P) & 0.121 & 0.211 & 0.308 & 0.410 & 0.597 & 0.845 & 0.983 & 0.496\\\hline
  \end{tabular}}
\label{tab:test4}
\end{table}

\subsection{Conclusion and Discussion}
We have proposed a novel approach for solving False-Positive Reduction tasks for lung nodule detection CAD systems, based on the representation of a nodule candidate as a set of points with known coordinates, radiodensity, and class probability, predicted by a detector. Such representation allows us to use a wide set of models designed to work with point cloud and graph data. Representation of lung tissue as a set of points is quite efficient: a major part of a lung's volume is air, and observation of it can be skipped without missing any important information. This leads us to much more lightweight models compared to traditional CNN 3D, including the usage of the entire patch extracted from a CT scan. 

We have provided an extensive comparison with SOTA results from open sources. For a FPR task we compared PointNet-like approaches with widely popular CNN-like models. We have shown that the DGCNN model can outperform CNN 3D at the False Positive reduction task. We have also provided an ablation study as essential for the paper and understanding of the method of the use of point clouds. The results show the importance of such parameters as radiology density ($HU$) and detector probability ($p$) in the context of presenting samples in the point cloud.

We have also presented augmentation techniques that lead to better model performance. According to these results, we assume that such a representation and approach can be successfully transferred to nodule detection tasks, and we plan to extend this work and construct the pipeline for nodule detection on point cloud representation on chest CT scans in its entirety. A major point that remains for further research will be the performance of tests on other datasets from different sources. 

\bibliographystyle{unsrt}  
\bibliography{arxiv2020}{}

\appendix

\section{ResNet3D details}
\label{app:app1}
\begin{figure}[htbp]
\includegraphics[width=1.\linewidth]{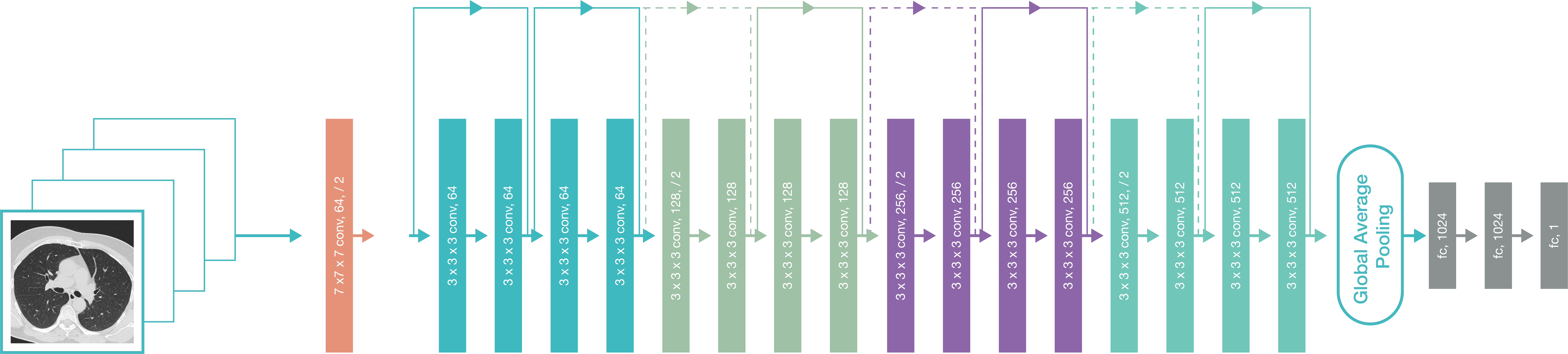}
\centering
\caption{Examples of point cloud samples for negative candidates}
\label{fig:resnet3d}
\end{figure}

We use modified ResNet18 as our baseline. We change all convolution layers to 3D convolutions and add two additional hidden fully connected layers after global average pooling with batch normalization and ReLU activation. As the final activation we use a sigmoid function (see details on Figure \ref{fig:resnet3d}). 
We use an ADAM optimizer with an initial learning rate equal to 0.001, train model during 75 epochs and decrease the learning rate twice every 10 epochs. We also add Gaussian blur, flips and random rotation as augmentation techniques. We balance the dataset by upsampling all positive candidates so positive and negative candidates ratios are equal to each other.

\section{Point Cloud samples visualisation}
\label{app:app2}
\begin{figure}[htbp]
\centering
\includegraphics[width=.4\linewidth]{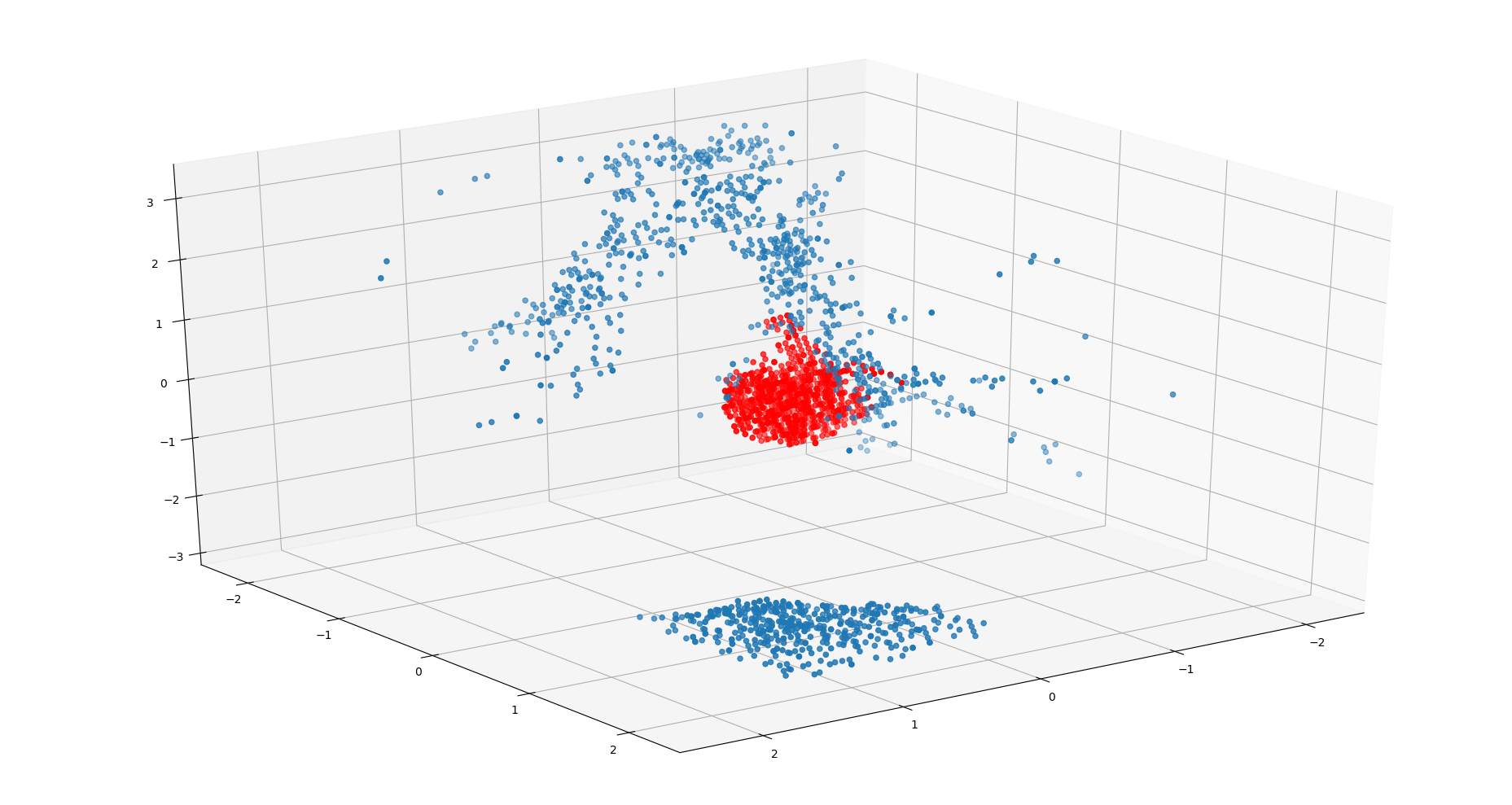}
\centering
\includegraphics[width=.4\linewidth]{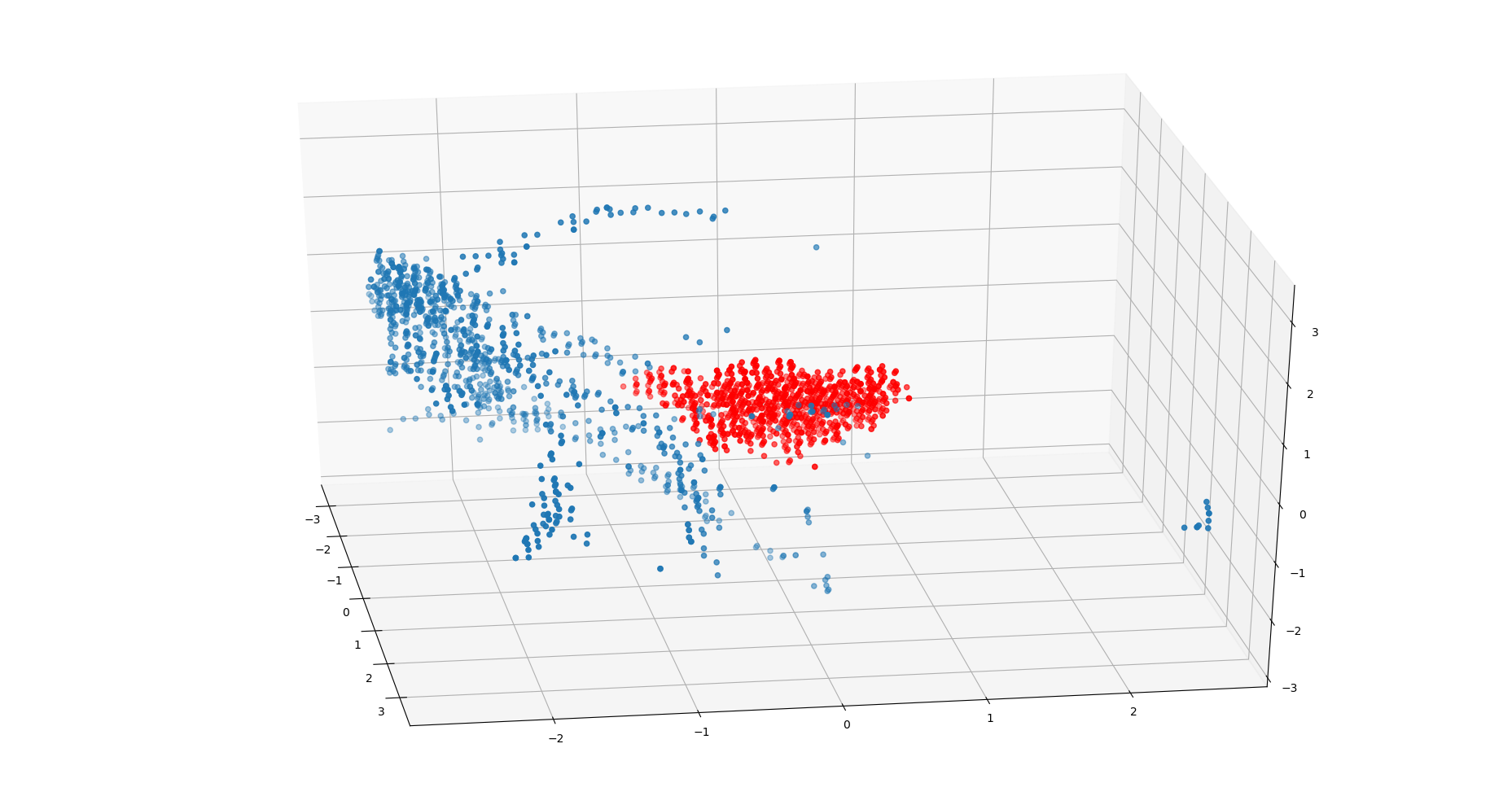}
\centering
\includegraphics[width=.4\linewidth]{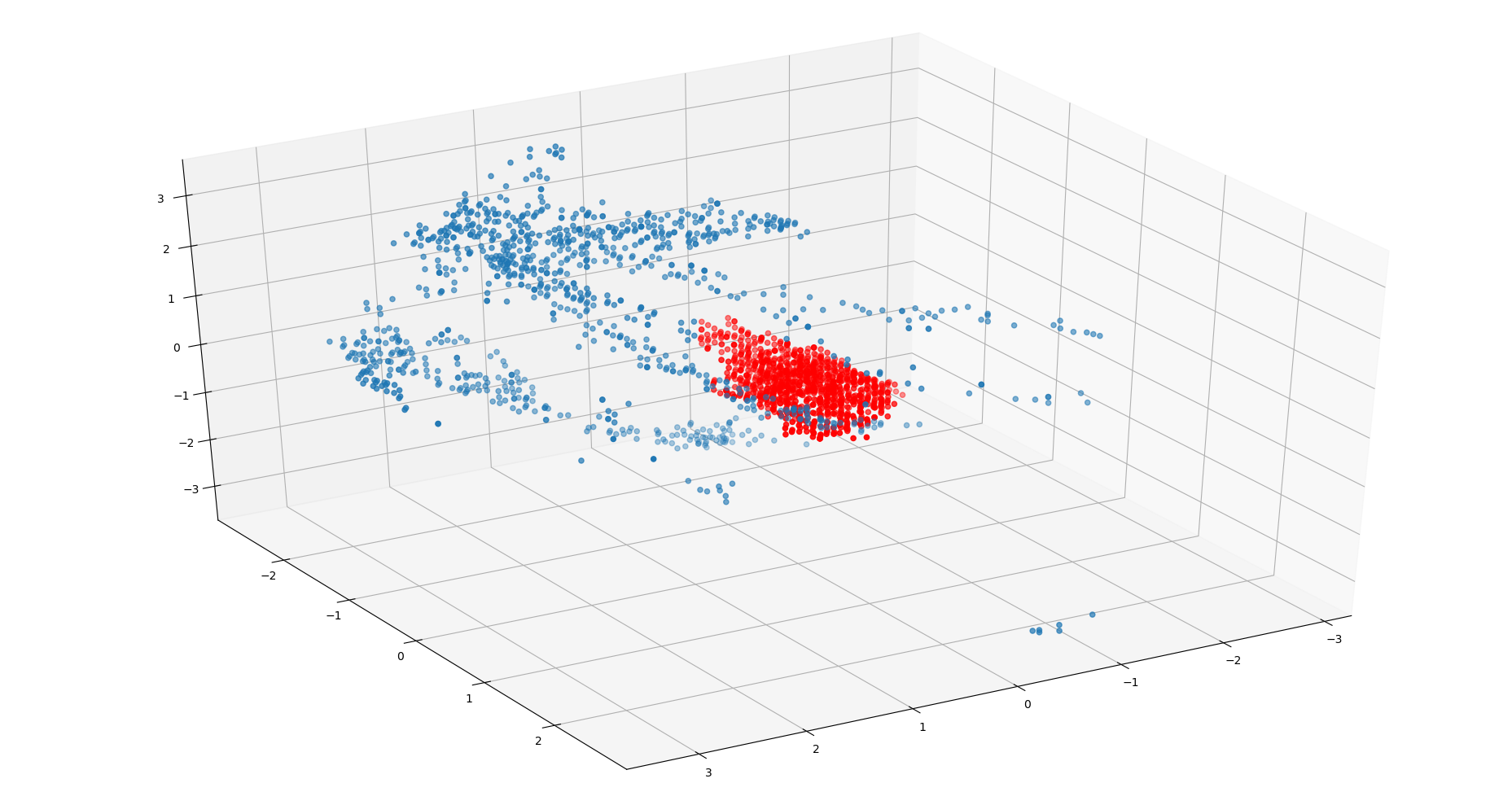}
\centering
\includegraphics[width=.4\linewidth]{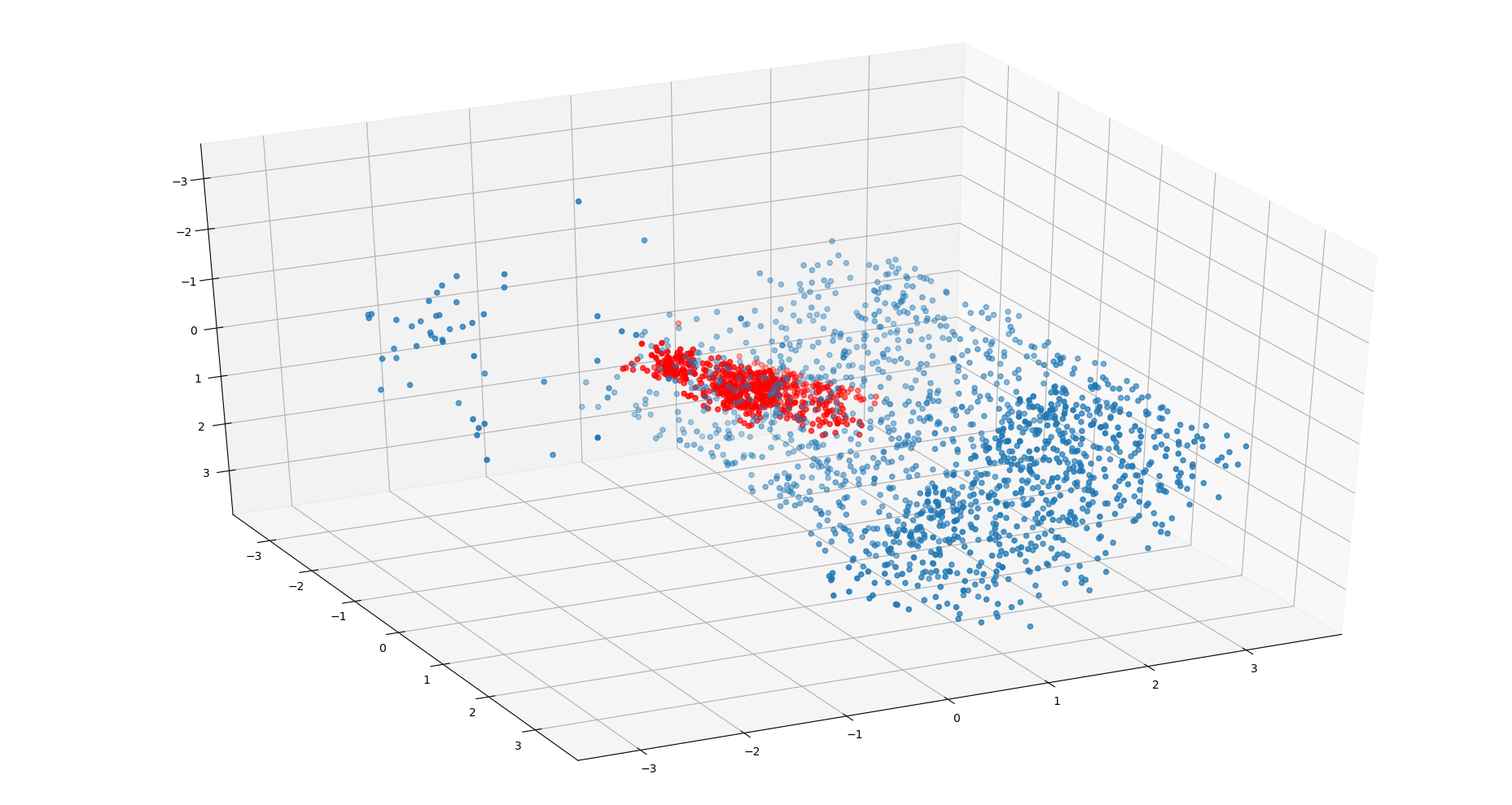}
\caption{Examples of point cloud samples for positive candidates}
\label{fig:pcsamples_pos}
\end{figure}
\begin{figure}[htbp]
\includegraphics[width=.4\linewidth]{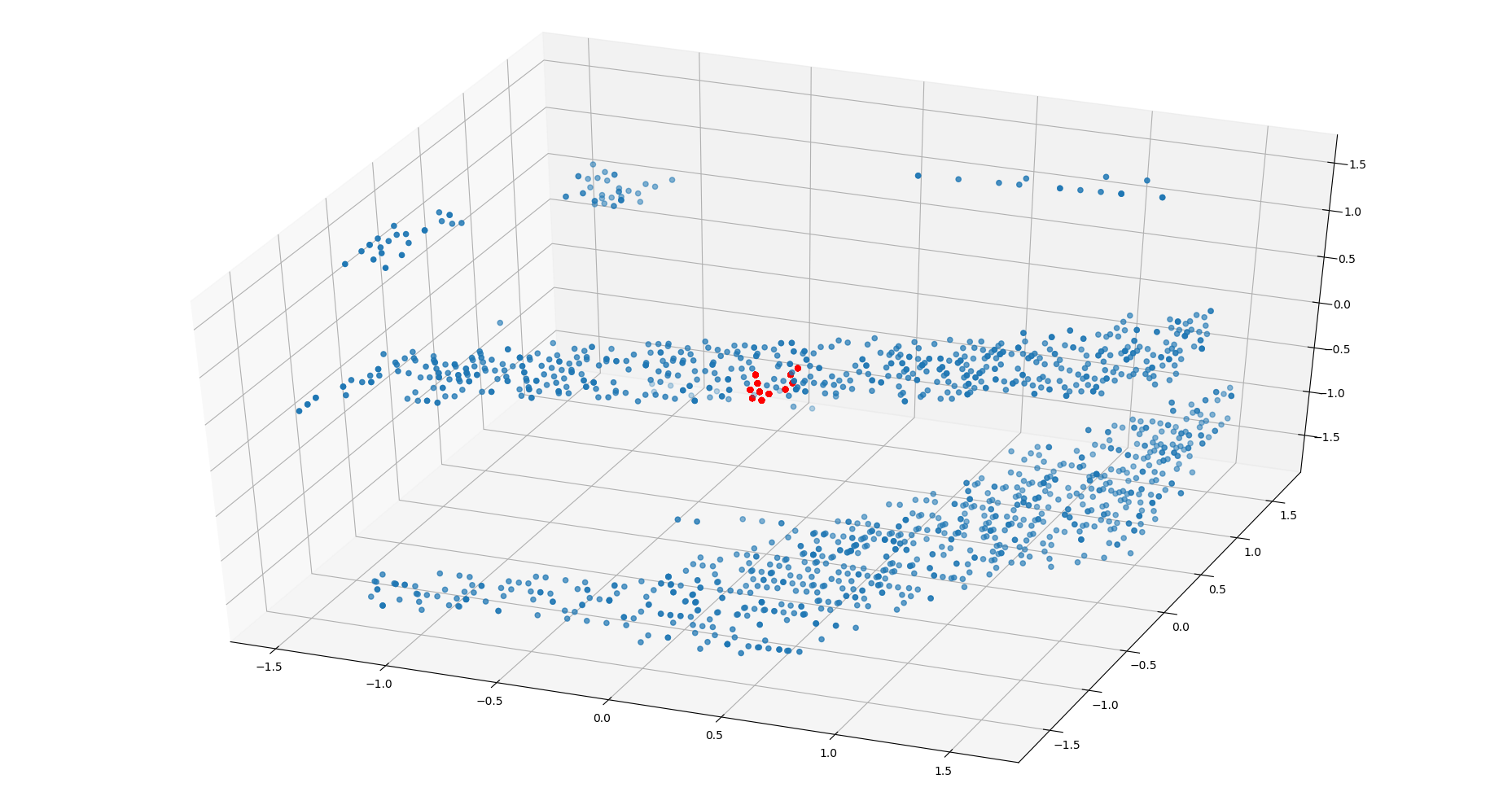}
\centering
\includegraphics[width=.4\linewidth]{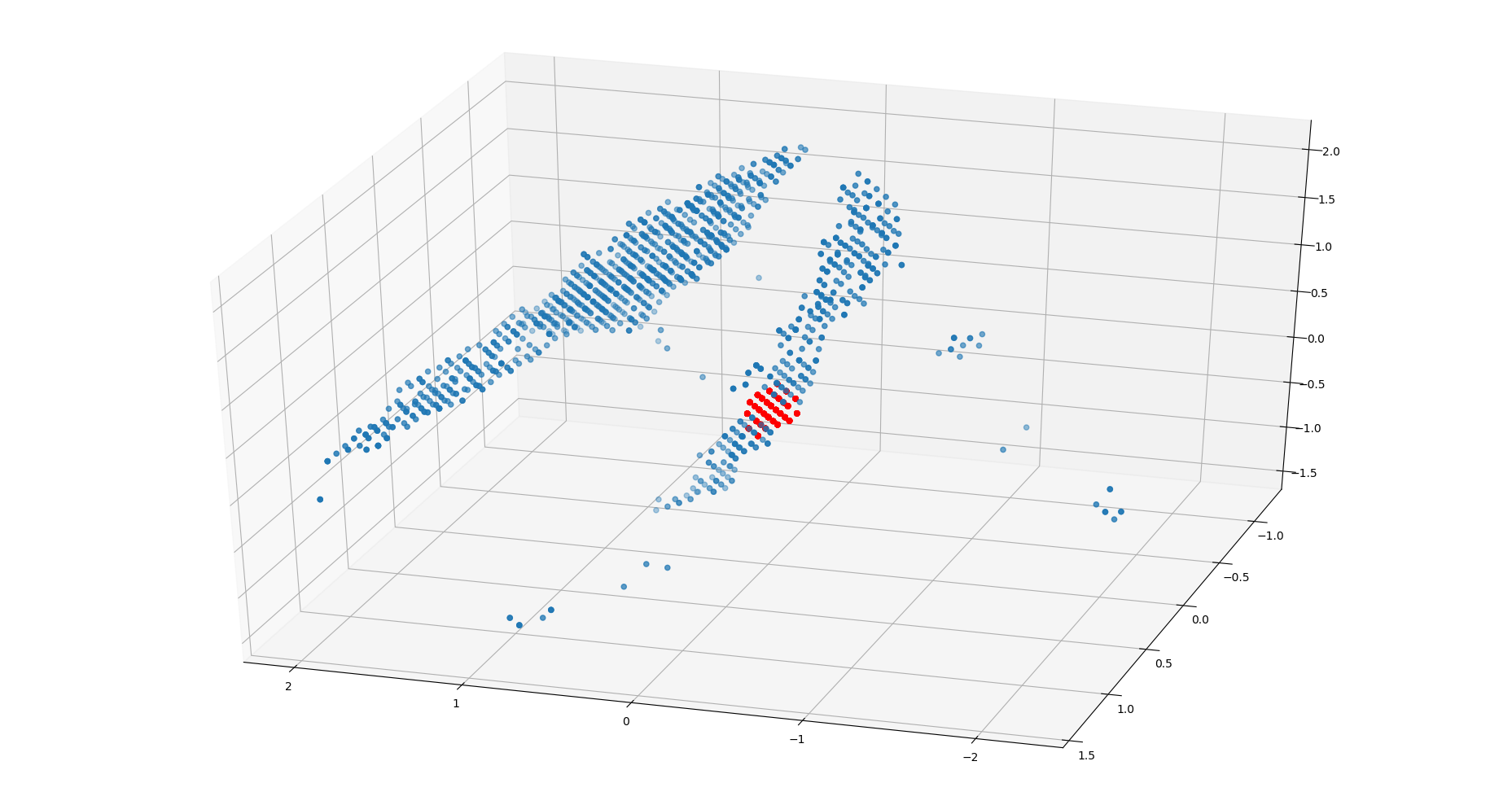}
\centering
\includegraphics[width=.4\linewidth]{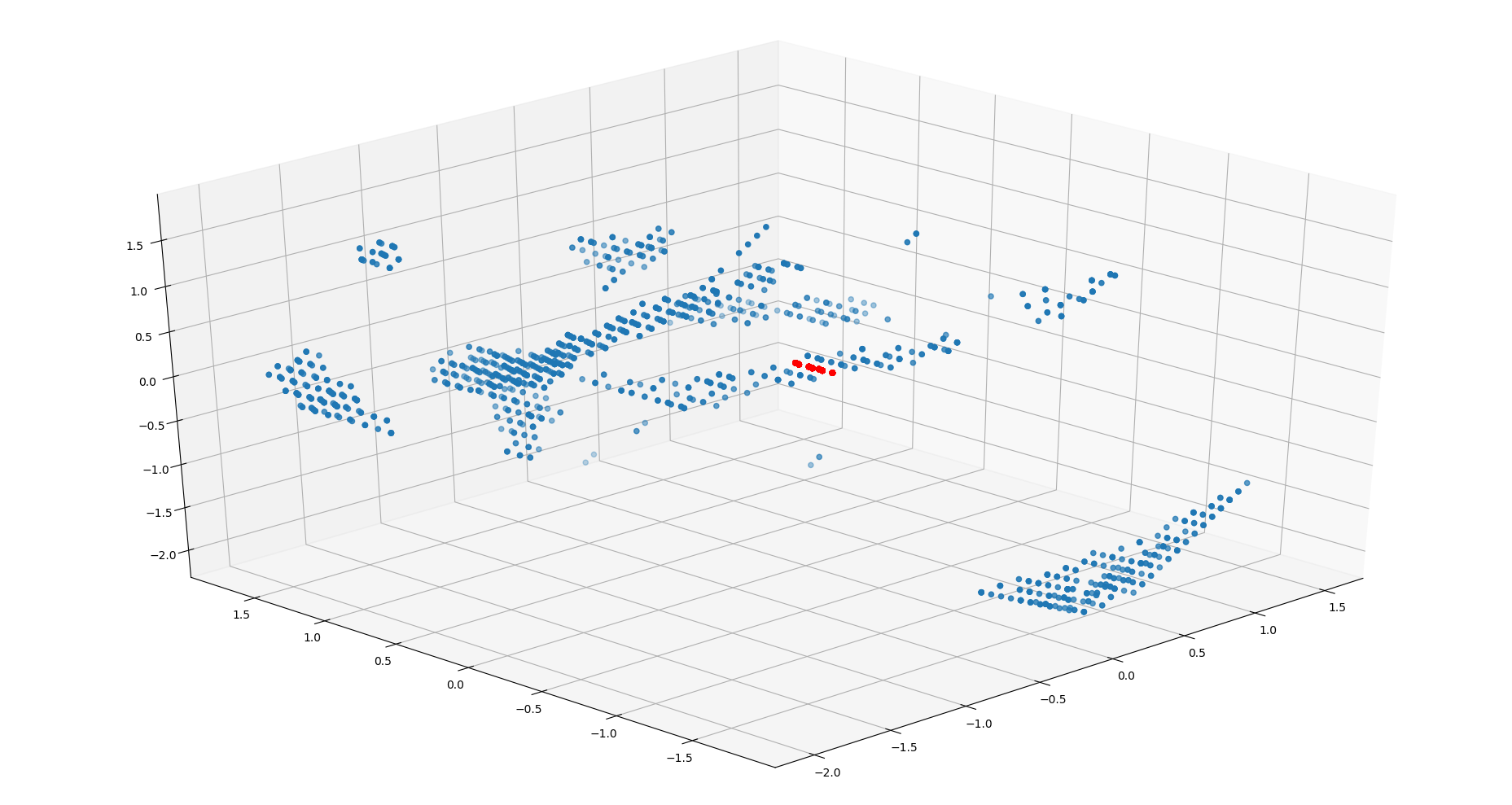}
\centering
\includegraphics[width=.4\linewidth]{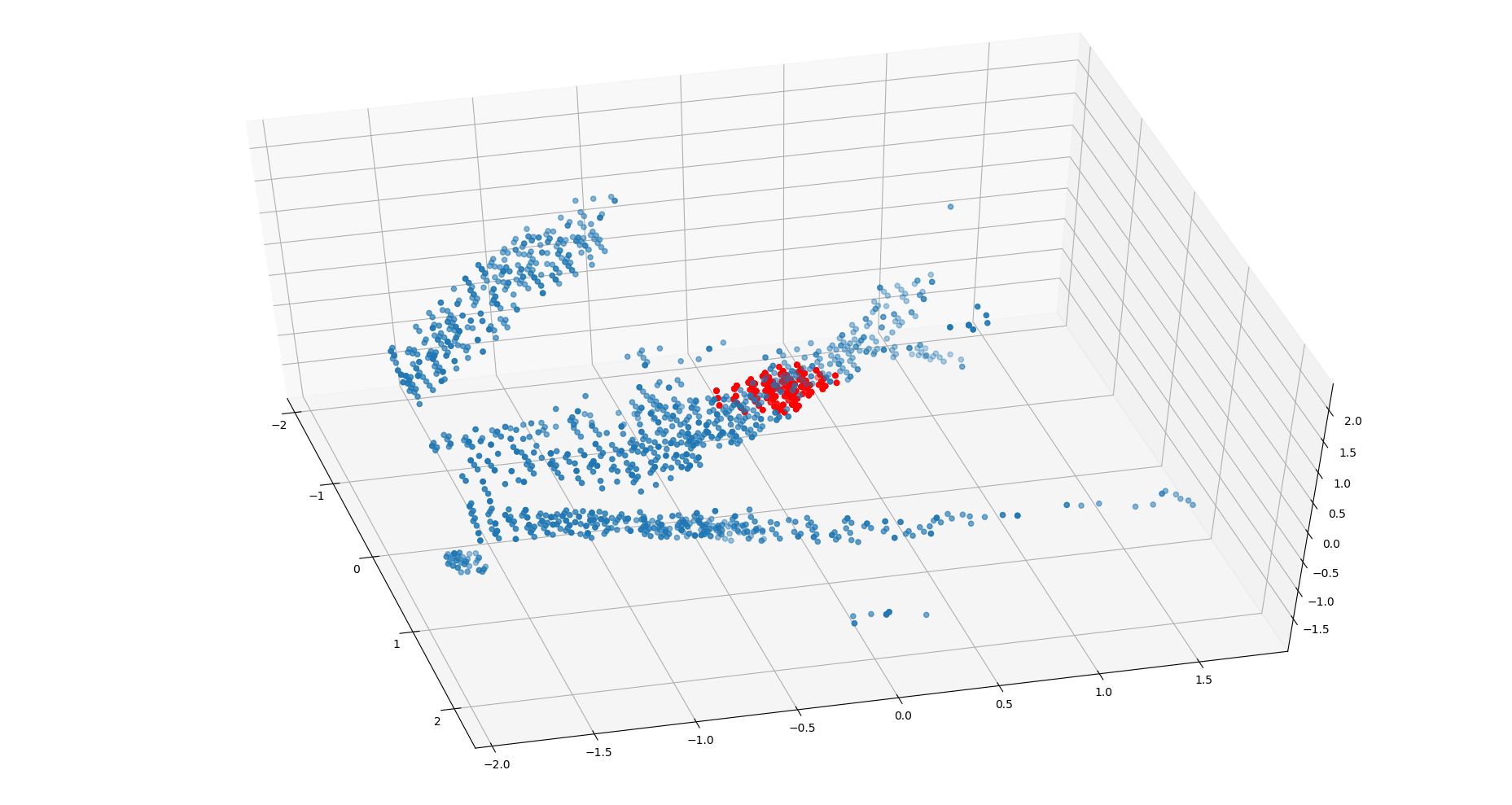}
\caption{Examples of point cloud samples for negative candidates}
\label{fig:pcsamples_neg}
\end{figure}

At Figure \ref{fig:pcsamples_pos} and Figure \ref{fig:pcsamples_neg} red points signify those points that detector marked as nodule candidates. Blue points represent the background. These plots show that point cloud contains all necessary information for successful separation of true positive candidates from false-positive candidates.  

\end{document}